\documentclass[5p,twocolum]{elsarticle}
\RequirePackage{lineno}
\usepackage{lineno,hyperref}

\journal{Astroparticle physics}

\begin{document}

\begin{frontmatter}

\title{Supernova Relic Neutrino Search with Neutron Tagging at Super-Kamiokande-IV}
\address[AFFicrr]{Kamioka Observatory, Institute for Cosmic Ray Research, University of Tokyo, Kamioka, Gifu 506-1205, Japan}
\address[AFFkashiwa]{Research Center for Cosmic Neutrinos, Institute for Cosmic Ray Research, University of Tokyo, Kashiwa, Chiba 277-8582, Japan}
\address[AFFipmu]{Kavli Institute for the Physics and
Mathematics of the Universe (WPI), Todai Institutes for Advanced Study,
University of Tokyo, Kashiwa, Chiba 277-8582, Japan }
\address[AFFmad]{Department of Theoretical Physics, University Autonoma Madrid, 28049 Madrid, Spain}
\address[AFFubc]{Department of Physics and Astronomy, University of British Columbia, Vancouver, BC, V6T1Z4, Canada}
\address[AFFbu]{Department of Physics, Boston University, Boston, MA 02215, USA}
\address[AFFbnl]{Physics Department, Brookhaven National Laboratory, Upton, NY 11973, USA}
\address[AFFuci]{Department of Physics and Astronomy, University of California, Irvine, Irvine, CA 92697-4575, USA }
\address[AFFcsu]{Department of Physics, California State University, Dominguez Hills, Carson, CA 90747, USA}
\address[AFFcnm]{Department of Physics, Chonnam National University, Kwangju 500-757, Korea}
\address[AFFduke]{Department of Physics, Duke University, Durham NC 27708, USA}
\address[AFFfukuoka]{Junior College, Fukuoka Institute of Technology, Fukuoka, Fukuoka 811-0295, Japan}
\address[AFFgmu]{Department of Physics, George Mason University, Fairfax, VA 22030, USA }
\address[AFFgifu]{Department of Physics, Gifu University, Gifu, Gifu 501-1193, Japan}
\address[AFFuh]{Department of Physics and Astronomy, University of Hawaii, Honolulu, HI 96822, USA}
\address[AFFkanagawa]{Physics Division, Department of Engineering, Kanagawa University, Kanagawa, Yokohama 221-8686, Japan}
\address[AFFkek]{High Energy Accelerator Research Organization (KEK), Tsukuba, Ibaraki 305-0801, Japan }
\address[AFFkobe]{Department of Physics, Kobe University, Kobe, Hyogo 657-8501, Japan}
\address[AFFkyoto]{Department of Physics, Kyoto University, Kyoto, Kyoto 606-8502, Japan}
\address[AFFumd]{Department of Physics, University of Maryland, College Park, MD 20742, USA }
\address[AFFmit]{Department of Physics, Massachusetts Institute of Technology, Cambridge, MA 02139, USA}
\address[AFFmiyagi]{Department of Physics, Miyagi University of Education, Sendai, Miyagi 980-0845, Japan}
\address[AFFnagoya]{Solar Terrestrial Environment Laboratory, Nagoya University, Nagoya, Aichi 464-8602, Japan}
\address[AFFpol]{National Centre For Nuclear Research, 00-681 Warsaw, Poland}
\address[AFFsuny]{Department of Physics and Astronomy, State University of New York at Stony Brook, NY 11794-3800, USA}
\address[AFFniigata]{Department of Physics, Niigata University, Niigata, Niigata 950-2181, Japan }
\address[AFFokayama]{Department of Physics, Okayama University, Okayama, Okayama 700-8530, Japan }
\address[AFFosaka]{Department of Physics, Osaka University, Toyonaka, Osaka 560-0043, Japan}
\address[AFFregina]{Department of Physics, University of Regina, 3737 Wascana Parkway, Regina, SK, S4SOA2, Canada}
\address[AFFseoul]{Department of Physics, Seoul National University, Seoul 151-742, Korea}
\address[AFFshizuokasc]{Department of Informatics in
Social Welfare, Shizuoka University of Welfare, Yaizu, Shizuoka, 425-8611, Japan}
\address[AFFskk]{Department of Physics, Sungkyunkwan University, Suwon 440-746, Korea}
\address[AFFtohoku]{Research Center for Neutrino Science, Tohoku University, Sendai, Miyagi 980-8578, Japan}
\address[AFFtokyo]{The University of Tokyo, Bunkyo, Tokyo 113-0033, Japan }
\address[AFFtoront]{Department of Physics, University of Toront, 60 St., Toront, Ontario, M5S1A7, Canada }
\address[AFFtriumf]{TRIUMF, 4004 Wesbrook Mall, Vancouver, BC, V6T2A3, Canada }
\address[AFFtokai]{Department of Physics, Tokai University, Hiratsuka, Kanagawa 259-1292, Japan}
\address[AFFtit]{Department of Physics, Tokyo Institute
for Technology, Meguro, Tokyo 152-8551, Japan }
\address[AFFtsinghua]{Department of Engineering Physics, Tsinghua University, Beijing, 100084, China}
\address[AFFwarsaw]{Institute of Experimental Physics, Warsaw University, 00-681 Warsaw, Poland }
\address[AFFuw]{Department of Physics, University of Washington, Seattle, WA 98195-1560, USA}

\author[AFFtsinghua]{H.~Zhang}
\author[AFFicrr,AFFipmu]{K.~Abe}
\author[AFFicrr,AFFipmu]{Y.~Hayato}
\author[AFFicrr]{K.~Iyogi}
\author[AFFicrr,AFFipmu]{J.~Kameda}
\author[AFFicrr,AFFipmu]{Y.~Kishimoto}
\author[AFFicrr,AFFipmu]{M.~Miura}
\author[AFFicrr,AFFipmu]{S.~Moriyama}
\author[AFFicrr,AFFipmu]{M.~Nakahata}
\author[AFFicrr]{Y.~Nakano}
% ...


\author[AFFuw]{K.~Connolly}
\author[AFFuw]{M.~Dziomba}
\author[AFFuw]{R.~J.~Wilkes}

\author{\\The Super-Kamiokande Collaboration}

\begin{abstract}
A search for Supernova Relic Neutrinos $\bar{\nu}_e$'s is first conducted 
via inverse-beta-decay by  
tagging neutron capture on hydrogen at Super-Kamiokande-IV. The neutron 
tagging efficiency is 
determined to be $(17.74\pm0.04_{stat.}\pm1.05_{sys.})\%$, while the 
corresponding accidental background  
probability is $(1.06\pm0.01_{stat.}\pm 0.18_{sys.})$\%. Using 960 days of 
data, we obtain 
13 inverse-beta-decay candidates in the range of $E_{\bar{\nu}_e}$ between 
13.3 MeV and 31.3 MeV. All of the observed candidates are attributed to
background. Upper limits at 90\% C.L. are calculated 
in the absence of a signal.
\end{abstract}

\begin{keyword}
\texttt Neutron tagging; Water Cherenkov detector; hydrogen; Supernova Relic Neutrinos 
\end{keyword}

\end{frontmatter}


\section{Introduction}
Neutrinos emitted from all past core-collapse
supernovae should form an isotropic flux. Sometimes called the 
Diffuse Supernova Neutrino Background (DSNB), these neutrinos
will be referred to as Supernova Relic Neutrinos (SRN) herein. Many models
have been constructed to predict the SRN flux and spectrum~\cite{Totani95,
Totani96, Malaney97, Hartmann97, Kaplinghat00, Strigari03, Ando03, Fukugita03, Lunardini09, HBD09, VP11, Nakazato13}.
Although all six types of neutrinos are emitted from a core-collapse supernova,
SRN's are most likely detected via the inverse beta decay (IBD)
reaction $\bar{\nu}_e p \rightarrow e^+ n$ in existing detectors.
Super-Kamiokande (SK) has previously carried out searches for SRNs from the 
expected IBD positrons without requiring the detection of a delayed neutron, 
placing an integral flux limit for $E_{\bar{\nu}_e}>17.3$ MeV 
($E_{\bar{\nu}_e}\approx E_{e^+}+1.3~\mathrm{MeV}$) 
in the absence of a signal~\cite{relic03,relic12}.
Since the detector cannot directly differentiate electrons from positrons 
(the positron annihilation signal is below Cherenkov threshold in water), 
these searches suffer from background of electrons and positrons. Some of 
these potential backgrounds include atmospheric neutrino $\nu_e/\bar{\nu}_e$ 
and $\nu_{\mu}/\bar{\nu}_{\mu}$ charged-current interactions and atmospheric 
neutrino neutral-current interactions. Many background channels either
do not produce neutrons or more than one neutron, but they can generate 
an electron or positron that passes all of the selection criteria, 
thereby contaminating the candidate samples.
Spallation backgrounds have limited the lower anti-neutrino energy 
threshold in previous SRN searches at SK. 

Positive identification of $\bar{\nu}_e$'s by tagging (and counting) 
neutrons in delayed coincidence will play a critical
role in both the suppression of backgrounds in those samples as well as in 
lowering the energy threshold. 
Kamland made the first attempt to search for the SRN flux down 
to 8.3 MeV by detecting IBDs with neutron capture on hydrogen 
in a one-kiloton liquid scintillator detector~\cite{kamland}. 
This paper will present a study to detect IBDs with neutron 
capture on hydrogen at SK, providing 
an improved search of SRNs from the previous threshold of 17.3 MeV down 
to the present 13.3 MeV, where greater SRN flux is expected. In addition,   
this study can be treated as an after-the-fact approach in parallel 
to the ongoing R$\&$D initiative aimed at   
detecting IBDs in water with enhanced neutron captures on dissolved gadolinium. 

\section{Experimental approaches to detect the neutron}
To detect the neutron signal, two independent approaches have been proposed 
to implement this capability in the SK experiment, a large underground water
Cherenkov detector containing 50 kilotons of pure water.
The detector consists of a cylindrical inner volume viewed by 11,129 
inward-facing 50-cm diameter photomultiplier tubes (PMTs),
surrounded by an outer annular volume viewed by 1,885 outward-facing 20-cm PMTs.
More detailed descriptions of the detector can be found 
elsewhere~\cite{skintroducion}. The first approach~\cite{gad} involves 
doping the water with a water-soluble chemical compound of gadolinium, 
neutron capture on which yields a gamma cascade with a total energy of 
about 8 MeV. These relatively high energy $\gamma$-rays should be readily 
seen by SK. The second approach is to detect the single 2.2 MeV $\gamma$ 
released from neutron capture on hydrogen~\cite{force}. This approach 
requires a 500-$\mu$s forced trigger scheme following a normal trigger, 
in order to identify the 2.2 MeV $\gamma$ offline.  

The detection of delayed-coincidence 2.2 MeV $\gamma$'s was first 
successfully demonstrated in SK using forced triggers~\cite{ntag}.
In the summer of 2008, SK's front-end electronics
were upgraded, after which began the data-taking period known as SK-IV. 
A major part of this upgrade was to use a 60 kHz periodic 
trigger to seamlessly read out
all PMTs all the time. In this data stream, events are searched for by
a software trigger, which is based on the number of coincident PMT hits
within 200 ns. To search for delayed-coincidence 2.2 MeV $\gamma$'s, a new
coincidence level was introduced: a super-high-energy (SHE) event requires
at least 70 coincident PMT hits corresponding to about 10 MeV (this level
was lowered to 58 or about 8 MeV in the summer of 2011). SHE events contain
all PMT hits from 5 $\mu$s before the SHE trigger time to 35
$\mu$s afterwards. The 5 $\mu$s of data prior the SHE trigger time provides a
chance to catch pre-activity events, e.g. a prompt $\gamma$-ray in a 
sub-Cherenkov muon background event produced by an
atmospheric $\nu_\mu$ interaction with oxygen. SHE events
without coincident outer detector activity are always followed 
by an after trigger
(AFT) which contains all PMT hits of the subsequent 500 $\mu$s.
The SK-IV data set used in this analysis was taken from November 22, 2008, to
December 27, 2011, with a total livetime of 960 days.

\section{The IBD event selection}
The IBD candidate search can be divided into two steps: one to find
the prompt signal with an energy
ranging from 12 to 30 MeV in the SHE trigger data; the other to tag the
IBD signal through the detection of a delayed event shortly after 
a prompt event is found.
The timing window for the delayed event ranges from 2 to 535 $\mu$s 
following the prompt event.
To avoid PMT signal reflection at the SK front-end electronics
after an event, the delayed event search starts
two $\mu$s after the prompt event time (defined by a GPS-synchronized clock).
The prompt events are selected by applying a number of cuts to suppress
muon-induced spallation background, atmospheric neutrinos,
solar neutrinos, and low energy radioactivities.
Details of the selection criteria for the prompt events, such as the
spallation cut, pre/post activity cut etc, can be found 
in~\cite{relic12}. Unlike the analysis in~\cite{relic12},
the reconstructed Cherenkov angle is required to be greater than 38
and less than 50 degrees. Also, the solar cosine angle cut of~\cite{relic12} is
loosened to 0.9 for energies below 16 MeV because of neutron tagging.
The number of remaining solar neutrino events in the sample is estimated 
to be about two events.
Table~\ref{tab:prompt} gives a summary of the event selection. Most 
efficiencies are evaluated using Monte Carlo (MC) simulation except for 
that of the spallation cut. The overall systematic uncertainty of the 
primary event
efficiency is $\sim$3.3\% due to uncertainties in the IBD cross
section (1.0\%), and the data reduction (3.1\% as given in~\cite{relic12}), which are added in quadrature.
\begin{table}
\caption{\label{tab:prompt}
Summary of the selection criteria for the prompt events ($N_{e^+}$) with energy
$E_{e^+}$ ranging from 12 to 30 MeV,  the number of events surviving each cut and 
the evaluated efficiency ($\epsilon_{e}$) from the large mixing angle 
model~\cite{Ando03}. Errors are statistical only.
}
\begin{tabular}{lrc}
Cuts                   & $N_{e^+}$ &$\epsilon_{e}$  (\%) \\ \hline
First reduction        & 49288    & 99.22$\pm$0.04\\
Spallation cut         &  2417    & 86.02$\pm$0.18\\
External event cut     &  2148    & 82.36$\pm$0.19\\
Solar cut              &  1649    & 81.54$\pm$0.20\\
Cherenkov angle cut    &   996    & 75.17$\pm$0.22\\
Pre/post activity cut  &   959    & 75.06$\pm$0.22\\
$\pi_{like}$ cut        &   948    & 74.32$\pm$0.22\\
Multi-ring cut         &   943    & 73.64$\pm$0.22\\
$\mu/\pi$ cut          &   942    & 73.14$\pm$0.23\\
\hline
\end{tabular}
\end{table}

An average 2.2 MeV $\gamma$ event in SK produces about seven recorded PMT hits, 
all of which share a common orientation in both time and space, but most of the 
hits have various arrival times due to the different travel distances. The flat 
timing distribution for these signal hits can be sharpened to form a timing peak 
if the common orientation or vertex is known. However, the continuous ~50 MHz dark 
noise and radioactive background for all the PMTs render a stand-alone reconstruction 
very difficult because of the long neutron lifetime. Fortunately, since neutrons
produced in IBD quickly thermalize and are eventually captured by
hydrogen with a mean free path length of $\sim$50 cm, to a 
good approximation the
location for an emission of a 2.2 MeV $\gamma$ can be treated to
share a common vertex with the prompt event, for
which the reconstructed position resolution is 40$\sim$50 cm.
The reconstructed vertex for the prompt event is therefore used to 
calculate the path length to each hit PMT, in order to
subtract the time-of-flight (TOF) from the measured light arrival time.
Due to a PMT timing resolution of 3 ns, the hits of real signal events cluster
within a 10 ns window, while hits due to PMT dark noise, radioactivity in the
surrounding rock, radon contamination events in water, and so on 
are typically more spread
out since the light does not originate from the primary event vertex. A 10 ns
sliding window is then applied to search for every 
timing peak and to give the number of PMT hits (N$_{10}$).
Fig.~\ref{fig:n10} shows the distribution of N$_{10}$ for a 
2.2 MeV $\gamma$ signal and background,
in which the signal events are from Monte Carlo simulation, while 
the background events are from the random trigger data. 

To remove background PMT hits the following selection criteria are applied:
\begin{figure}[!t]
   \centering
  \includegraphics[width=1.0\columnwidth]{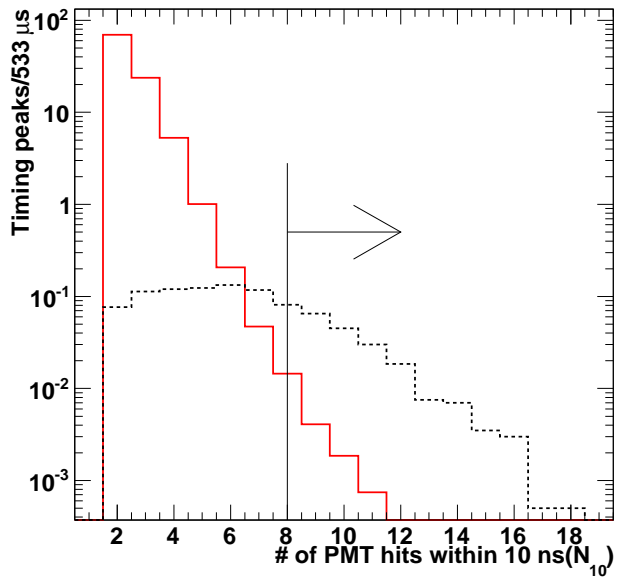}
  \renewcommand{\figurename}{Fig.}
   \caption{Distribution of N$_{10}$ for 2.2 MeV $\gamma$ from MC 
(dashed) and background (solid) from random trigger data. The arrow indicates the cut for selecting the signal.}
   \label{fig:n10}
 \end{figure}
The number of PMT hits in N$_{10}$ is required to be greater than 7. 
It is noted that about 86\% of
low energy $\gamma$ background events originating near the detector wall have 
clustered PMTs in both space and time, and thus survive the N$_{10}$ cut. 
The number of PMT hits in clusters (N$_{\mbox{cluster}}$) is defined for 
those within 10 ns and less than $14.1^\circ$ as seen from the positron 
candidate vertex. A cut on the quantity of N$_{10} - $N$_{\mbox{cluster}}$ is 
required to be greater than 5.
Since most of the light of a 2.2 MeV gamma originates from
a single Compton electron, most hits are in the same detector
hemisphere, while PMT noise is distributed more uniformly. As a result, a 
summed vector is
calculated for all the hits in N$_{10}$. Angles between the individual vector
for each hit and the summed vector are computed, giving the number of
hits with an angle greater than $90^\circ$ (N$_{\mbox{back}}$). The delayed
event should be  N$_{10}-$N$_{\mbox{back}}>$ 6, which removes
events with many background hits in the backward hemisphere. Some PMTs
are more likely to be illuminated than others for a given vertex.
A hit probability of the i-th PMT is defined by
$\frac{(\cos \theta_i)_{eff}}{R_i^2}e^{-R_i/L} $, where $R_i$ is the
distance from the vertex to the i-th PMT, $\theta_i$ is the incident
angle, $(\cos \theta_i)_{eff}$ includes the angular dependence of PMT
geometry and reflection/absorption of acrylic case, and $L$ is the attenuation 
length of Cherenkov light in the SK water. The calculated
hit probabilities for all the hits are sorted in decreasing order. The number
of hits accounting for the bottom 25-50\% of the summed probabilities is
defined as N$_{\mbox{low}}$. The fraction varies with the vertex location
and is set to 50\% when the vertex is close to the wall.
A cut on  N$_{10} - $N$_{\mbox{low}}$ is required to be greater than 4. 
The final reduction utilizes
a likelihood ratio based on four discriminating variables: number of
PMT hits within $\pm$150 ns around N$_{10}$ peak, root mean square of 
the N$_{10}$ timing peak,
root mean square of the azimuth angles for all the PMT hit vectors along the
summed vector, and mean value of opening angle between the PMT hit vectors and
the summed vector. The likelihood ratio is
required to be greater than 0.35. 

Table~\ref{tab:delayed} gives a
summary for the selection criteria, the background probability, and 
the efficiency of the delayed events for each cut.
The efficiency of the delayed event is corrected with a 
factor of $\sim$92\% due to the 
533 $\mu$s time window. It is observed that
most delayed  events cannot fire sufficient PMTs to meet the
minimum requirement on N$_{10}$.
\begin{table}
\caption{\label{tab:delayed}
Summary of the selection criteria,  the
probability of accidental background, and the efficiency of finding 
delayed events. The samples are from the
random trigger (background) and Monte Carlo simulation 
(2.2 MeV $\gamma$ signal).
All efficiencies for the delayed events are corrected 
by $\sim$92\% due to the width of the 533 $\mu$s time window.
Errors are statistical only.  See text for variable name definitions.}
\begin{tabular}{lrc}
Cuts                   &  Bkg Prob. (\%)                        & Efficiency(\%) \\ \hline
N$_{10}>$ 7                &           100                        & 30.19$\pm$0.04\\
N$_{10}-$N$_{\mbox{cluster}}>$ 5   &  25.48$\pm$0.04                        & 28.27$\pm$0.04\\
N$_{10} - $N$_{\mbox{back}}>$ 6       &  21.13$\pm$0.04                        & 26.78$\pm$0.04\\
N$_{10}- $N$_{\mbox{low}}>$ 4       &   4.14$\pm$0.02                        & 19.11$\pm$0.04\\
Likelihood ratio $>$ 0.35 &   1.06$\pm$0.01                        & 17.74$\pm$0.04\\
\hline
\end{tabular}
\end{table}
Basing the TOF correction on the SRN candidate vertex 
(rather than the true vertex of 
the delayed event) changes the efficiency of finding
delayed events by at most 2.5\% relatively. Uniformity of both the 
MC signal efficiency and the
background probability were studied using 110 positions within the detector.
These spatial variations of MC signal efficiency and
background probability were found to be 5.9\% and 16.8\%, respectively.
These variations were then assigned to the systematic
uncertainties. Therefore, the efficiency and the background probability 
for the delayed events are
$(17.74\pm0.04_{stat.}\pm1.05_{sys.})\%$ and 
$(1.06\pm0.01_{stat.}\pm 0.18_{sys.})$\%, respectively.
Combining the primary event efficiency and delayed event efficiency, 
the IBD detection efficiency
($\epsilon$) is obtained to be $(13.0\pm0.8)$\%.

\section{Test with Am/Be source data}
\begin{figure}[!t]
   \centering
  \includegraphics[width=1.0\columnwidth]{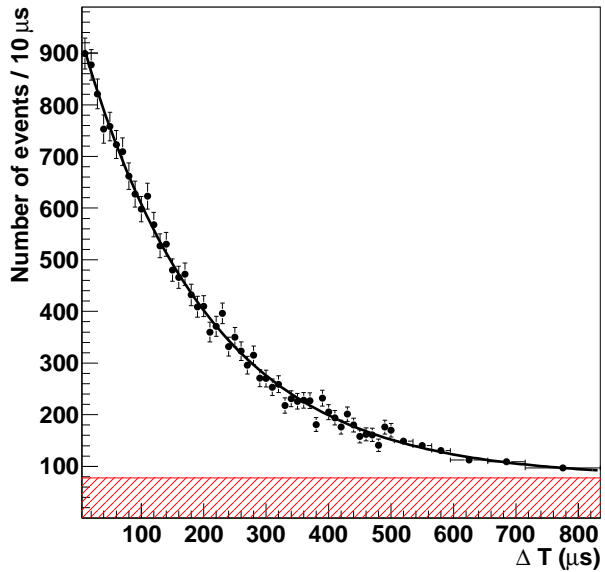}
  \renewcommand{\figurename}{Fig.}
   \caption{Distribution of $\Delta \textrm{T}$ for the Am/Be data(points). 
   The curve is for the fitting results. The shaded histogram indicates 
the expected background. Errors are statistical only.}
   \label{fig:dtc}
 \end{figure}

To verify the detection efficiency for the 2.2 MeV $\gamma$'s given in 
Table~\ref{tab:delayed}, a test was
carried out using an Am/Be source embedded in a 
bismuth germanate (BGO) scintillator during SK-IV. The experimental setup
and other details can be found elsewhere~\cite{ntag}.
The experimental apparatus was deployed at certain positions in the SK tank, 
during which the forced trigger gate for catching the 2.2 MeV $\gamma$'s
was temporarily enlarged to 800 $\mu$s in order to obtain a more 
complete neutron lifetime spectrum.
To get the time distribution of the source-related background and 
accidental background, 10 Hz of
800$\mu$s random trigger data was also taken. Fig.~\ref{fig:dtc}
shows the distribution of time differences ($\Delta$T) between the 
delayed events and the prompt events, which is fitted with an 
exponential plus a constant with the signal fraction as a free parameter, 
to give the neutron lifetime in water.
In order to verify the neutron lifetime time and examine possible 
position dependence of detection efficiency,
the source was deployed at three different locations: at the 
center of the tank, close to the wall, and close to the top.
All of the resulting lifetime measurements were
consistent within one standard deviation. The average neutron lifetime in water was
found to be ($203.7\pm2.8$)$\mu$s. The efficiencies measured at 
the three locations are
in agreement within 10\%, which also agrees with the estimation 
of Monte Carlo simulation. The average efficiency of
$(19.0\pm0.2)$\% in this enlarged 800 $\mu$s window is in good 
agreement with the value of $(19.2\pm0.1)$\% estimated from MC simulation.

\section{Analysis and Results}
\begin{figure}[!t]
   \centering
  \includegraphics[width=1.0\columnwidth]{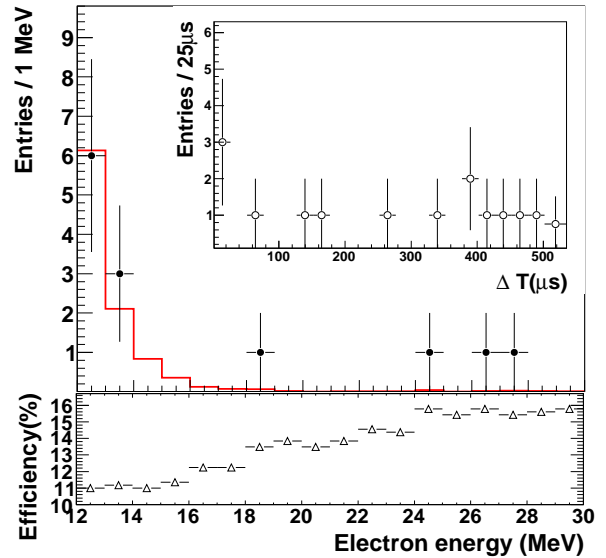}
  \renewcommand{\figurename}{Fig.}
   \caption{Positron energy spectrum of the IBD candidates (points). 
The histogram represents the expected accidental background. The plot 
embedded in the upper right shows the timing offset for the delayed 
candidates. Shown at the bottom of the figure is a plot of the IBD 
detection efficiency for each energy bin; the jumps at 18 MeV and 24 MeV 
are due to energy-dependent spallation cuts. Errors are statistical only.}
   \label{fig:electron}
 \end{figure}

Returning to the low energy SRN search using 960 live days of SK-IV data, 
after passing the selection criteria for both the prompt events and the delayed 
events the relevant distributions for the remaining events with at least 
one neutron candidate are shown in Fig.~\ref{fig:electron}.
There are 13 IBD candidates observed consistent with accidental 
background events evaluated to be $10\pm1.7$.
Two out of these 13 primary events with electron energies around 12 MeV 
are observed to have two neutron candidates, which indicates they 
are likely to be from spallation backgrounds with high neutron multiplicity. 

A number of studies have been performed to
provide insight into the origin of possible background in this 
energy domain, especially those arising from
atmospheric $\bar{\nu}_{\mu}/\nu_{\mu}$ CC interaction, $\pi^{\pm}$ 
production and NC interactions with water. This is achieved by changing 
the cut on the Cherenkov
angle $\theta_C$ of the primary event, in which an electron/positron 
is defined with $38^\circ <\theta_C<50^\circ$.
The $\mu^{\pm}$ and $\pi^{\pm}$ events are defined with $\theta_C < 38^\circ$,
while the NC events are defined with $\theta_C > 50^\circ$. There are 
144 primary $\mu^{\pm}$ and $\pi^{\pm}$ candidate
events with 22 delayed candidates
and 489 NC candidate events with 47 delayed candidates.
A clear neutron lifetime curve is observed in both delayed 
candidate samples, showing that the primary events are indeed accompanied by
neutrons. The flat timing offset distribution for the delayed candidates 
in Fig.~\ref{fig:electron} does not show significant leakage from these 
two types of physical backgrounds. The number of atmospheric 
$\bar{\nu}_e$ events is estimated to be 0.1, 
while the number of the $\bar{\nu}_\mu$ events is about 1.0. 
The later is due to the  $\bar{\nu}_\mu$ charged-current interaction, 
which produces a delayed neutron and a positron from an invisible $\mu^+$ Michel decay. 
In absence of a significant 
signal the Rolke method~\cite{rolke} is used to convert the 
number of observed and expected background events
$n_{obs}=13$ and $n_{bkg}=10.0\pm1.7$ to a 90\% C.L upper 
limit of 80.1 events in total 
or 30.5 events/22.5~(kton$\cdot$year), 
taking into into account 
the IBD detection efficiency $\epsilon$.
Table~\ref{tab:model} lists the expected number of SRN events in 22.5
kton$\cdot$year for different models. The upper
limit on the SRN flux $F_{90}$ can be derived from $N^{\prime}_{90}$ 
using the following simple relation:
\begin{equation}
 F_{90}=\frac{N^{\prime}_{90}}{N_P} \times F_M
 \end{equation}
where $F_M$(cm$^{-2}$s$^{-1}$) is the total flux for a certain model 
and $N_P$ is the
predicted annual event rate in the energy range which can be found in 
Table~\ref{tab:model}.
This table also contains upper limits ($F_{90}$) at 90\% C.L. for 
different models and the
predicted annual event rate ($T_P$) after efficiency correction.
\begin{table}
\caption{\label{tab:model}Total flux for each SRN model ($F_M$), 
predicted number
of SRN events in 22.5 kton$\cdot$year
with a neutrino energy range of 13.3$\sim$31.3 MeV ($N_P$), predicted number
of SRN events in 22.5 kton$\cdot$year
with a neutrino energy range of 13.3$\sim$31.3 MeV ($T_P$) after IBD 
efficiency correction and 
flux upper limit at 90\% C.L. ($F_{90}$)(cm$^{-2}$s$^{-1}$).}
\begin{tabular}{lcccc}
 SRN model & $F_M$ &  $N_P$ & $T_P$ & $F_{90}$\\
\hline
Constant SN \cite{Totani95} & 52.3 & 10.8 & 1.4 & 147.5 \\
HBD 6 MeV \cite{HBD09} & 21.8 & 4.4 & 0.6 & 150.9 \\
Chemical evolution \cite{Hartmann97}& 8.5 & 1.5 & 0.2 & 172.6\\
Heavy metal \cite{Kaplinghat00, Strigari03} & 31.3 & 4.7 & 0.6 & 201.8\\
LMA \cite{Ando03} & 28.8& 4.2 & 0.5 & 208.8\\
Failed SN \cite{Lunardini09} & 12.0 & 1.7 & 0.2 & 214.9\\
Cosmic gas \cite{Malaney97} & 5.3 & 0.7 & 0.1 & 230.6\\
Star formation rate \cite{Fukugita03} & 18.7 & 1.8 & 0.2 & 316.3 \\
Population synthesis \cite{Totani96} & 42.1 & 1.3 & 0.2 & 986.1\\
\hline
\end{tabular}
\end{table}

\begin{figure}[!t]
   \centering
  \includegraphics[width=1.0\columnwidth]{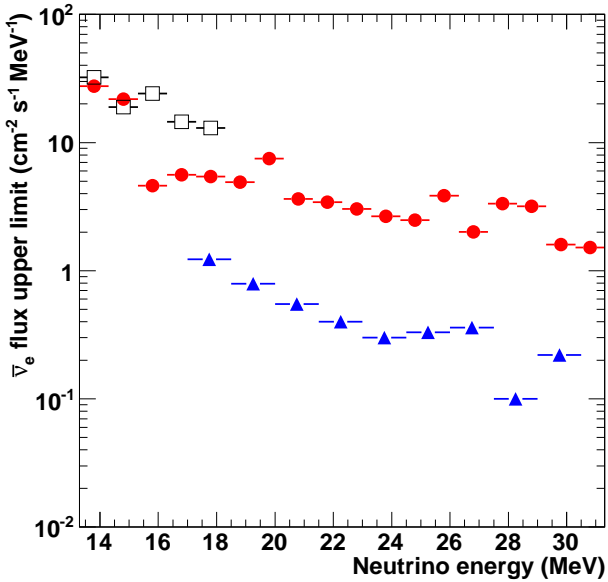}
  \renewcommand{\figurename}{Fig.}
   \caption{Model-independent 90\% C.L. differential upper limits on 
SRN $\bar{\nu}_e$ for SK-IV (solid circle).
For comparison, both KamLAND result (open square)~\cite{kamland}
and previous SK result (solid triangle) are also shown.}
   \label{fig:uplimit}
 \end{figure}

Model-independent $\bar{\nu}_e$ differential flux upper limits with 
one MeV energy bins are also calculated.
The 90\% C.L upper limits are calculated by
\begin{equation}
\phi_{90}=\frac{N_{90}}{T\cdot N_p \cdot \bar{\sigma}}
\end{equation}
where $N_{90}$ is the upper limit at 90\% C.L. in each energy bin, $T$ 
is livetime in seconds, $N_p$ is the number of free protons, 
$\bar{\sigma}$ is the average cross section for IBD at the center of each 
energy bin, and $\epsilon$ is the 
IBD detection efficiency for each energy bin.
Fig.~\ref{fig:uplimit} shows the upper limits for $\bar{\nu}_e$ 
in the energy range of 13.3$\sim$31.3 MeV.
Limits from KamLAND~\cite{kamland} based on 2343 live-days are 
also shown for comparison.

The previous SK search for SRN IBD positrons in \cite{relic12}
placed an integral 90\% C.L. limit on the SRN flux above 17.3 MeV 
neutrino energy of 2.9 cm$^{-2}$s$^{-1}$
(LMA model~\cite{Ando03}). In that search, SRN signal and
atmospheric neutrino backgrounds were
fitted to the energy spectra of the data for three different samples
differentiated by the reconstructed Cherenkov angle with an extended 
unbinned maximum likelihood
method. The SRN signal populates the single electron-like sample
($38^\circ <\theta_C<50^\circ$) below 30 MeV (signal region). Various
types of atmospheric neutrino background dominate the
background region above 30 MeV as well as the other
two (background) samples. 
To compare with the SK-IV differential limits in this paper, the previous SK 
background spectra as well as the SRN candidate positron spectra 
above 30 MeV (total energy) were fit to only atmospheric neutrino 
background contributions. The resulting background fit was
extrapolated in the signal region between 16 and 29.5 MeV (total positron 
energy) taking into account statistical
and systematic uncertainties. The data was divided into nine bins of 
1.5 MeV. Fig.~\ref{fig:uplimit} shows the 90\% C.L.
upper flux limits derived for each bin based on the background 
expectations with Gaussian uncertainties and
the IBD cross section evaluated at the bin center.
Below 17.3 MeV spallation background increases exponentially, so SRN detection
in that energy range is very difficult without neutron tagging. As the
SRN flux per MeV rises with decreasing energy, the region below 17.3
MeV is the most sensitive.

Since this study covers the high end of the  solar neutrino spectrum,  
a solar $\bar{\nu}_e$ upper limit at 90\% C.L. of the annual event rate 
is also calculated, giving an estimate of 21.2 events/22.5~kton$\cdot$year.
This corresponds to $4.2\times10^{-4}\times F_{SSM}$,
where $F_{SSM}$ is the solar $\nu_e$ flux predicted by the Standard Solar 
Model~\cite{SSM}.
This limit is 20 times more stringent than the previous SK result~\cite{SKI} 
due to the powerful background reduction provided by neutron tagging. However,  
note that the limit is an order less stringent than the KamLAND 
result~\cite{kamland} because of the higher neutrino energy threshold. 

\section{Summary and outlook}
In summary, a search for SRN $\bar{\nu}_e$ at SK-IV is first conducted 
via IBDs by tagging neutron capture on hydrogen. The neutron tagging 
efficiency is determined to be 
$(17.74\pm0.04_{stat.}\pm1.05_{sys.})\%$, while the corresponding accidental 
background probability is $(1.06\pm0.01_{stat.}\pm 0.18_{sys.})$\%. No 
appreciable IBD signal in the distribution of neutron lifetime is found 
using 960 days of data. The number of observed IBD candidates are 
consistent with the expected accidental background. A model-independent 
differential flux upper limit at SK is first derived from the previous 17.3 MeV 
threshold down to 13.3 MeV of the electron anti-neutrino energy.  

With more data collected and after further efforts in suppressing spallation 
background, it is expected that the neutrino energy threshold can be 
lowered down to 10 MeV and the better SRN flux limit can eventually 
be obtained with neutron capture on hydrogen at SK-IV. 
In addition, intense R$\&$D is currently underway 
towards a gadolinium-enhanced SK. The higher signal detection efficiency 
and greater background rejection provided by neutron capture on gadolinium, 
as well as the lowered energy threshold it makes possible, are expected --  
in the not-too-distant future -- to greatly improve SK$^\prime$s sensitivity 
and ultimately provide the world's first observation of the SRN signal.  

\section{Acknowledgment}

The authors gratefully acknowledge the cooperation of the Kamioka 
Mining and Smelting Company. The Super-Kamiokande detector was built and
operated with funds provided by the Japanese Ministry of Education, 
Culture, Sports and Technology, the United
States Department of Energy, and the U.S. National
Science Foundation. This work is also supported by
the National Natural Science Foundation of China (Grants No.11235006).

\section{References}


\end{document}